\documentclass[%
reprint,
superscriptaddress,
showpacs,
 amsmath,amssymb,
 aps,
 prl,
]{revtex4-2}
\usepackage{graphicx}
\usepackage{dcolumn}
\usepackage{bm}
\usepackage{hyperref}
\usepackage[mathlines]{lineno}
\usepackage{amsmath}
\usepackage{xcolor}
\usepackage{csquotes}
\pagecolor{white}
\begin{document}
\title{Narrowband THz Emission from a Plasma Oscillator Imbedded in a Plasma Density Gradient}
\author{Manoj Kumar}
\email[]{manojailum@gmail.com}
\affiliation{Department of Physics, Ulsan National Institute of Science and Technology, \\Ulsan 44919, Republic of Korea}
\author{Bernhard Ersfeld}
\affiliation{Department of Physics, Scottish Universities Physics Alliance and University of Strathclyde, \\Glasgow G4 0NG, United Kingdom}
\author{Jaeho Lee}
\affiliation{Department of Physics, Ulsan National Institute of Science and Technology, \\Ulsan 44919, Republic of Korea}
\author{Dohyun Park}
\affiliation{Department of Physics, Ulsan National Institute of Science and Technology, \\Ulsan 44919, Republic of Korea}
\author{Seungyun Kim}
\affiliation{Department of Physics, Ulsan National Institute of Science and Technology, \\Ulsan 44919, Republic of Korea}
\author{Inhyuk Nam}
\affiliation{Pohang Accelerator Laboratory, Pohang University of Science and Technology, \\Pohang 37673, Republic of Korea}
\author{Minseok Kim}
\affiliation{Pohang Accelerator Laboratory, Pohang University of Science and Technology, \\Pohang 37673, Republic of Korea}
\author{Seongjin Jeon}
\affiliation{Department of Physics and Photon Science, Gwangju Institute of Science and Technology, \\Gwangju 61005, Republic of Korea}
\author{Dino A. Jaroszynski}
\email[]{d.a.jaroszynski@strath.ac.uk}
\affiliation{Department of Physics, Scottish Universities Physics Alliance and University of Strathclyde, \\Glasgow G4 0NG, United Kingdom}
\author{Hyyong Suk}
\email[]{hysuk@gist.ac.kr}
\affiliation{Department of Physics and Photon Science, Gwangju Institute of Science and Technology, \\Gwangju 61005, Republic of Korea}
\author{Min Sup Hur}
\email[]{mshur@unist.ac.kr}
\affiliation{Department of Physics, Ulsan National Institute of Science and Technology, \\Ulsan 44919, Republic of Korea}


\begin{abstract}
A novel method is presented for generating radiation using the beat wave associated with a bi-frequency laser pulse, to excite plasma oscillations in a plasma slab with a density gradient. By resonantly exciting a plasma wave, it can be localised and transformed into a plasma oscillator that produces a beam of radially polarised terahertz radiation. Particle-in-cell simulations and analytic theory are used to demonstrate its main characteristics, which includes narrow bandwidth. The radiator should have useful applications such as terahertz-band particle accelerators and pump-probe experiments. 
\end{abstract}

\maketitle

There are numerous technologies that exploit terahertz (THz) electromagnetic waves. Significant strides have recently been made in utilising THz radiation \cite{Pickwell1, MM2, SP3, KG4, Kemp5}, in applications including THz particle accelerators \cite{Nanni6, Zhang7}, the excitation of material transitions \cite{Beck8}, and undertaking pump-probe experiments to investigate molecular bonding \cite{Forst9, Bakker10}. Most of these require narrow bandwidth THz pulses with GV/m-level field strengths. Current solid-state THz devices are limited by crystal and material damage at high intensities \cite{Fulop11, Zhang12}, and the free-electron laser \cite{VK13},  albeit powerful, is only available at large and expensive facilities. Plasma, on the other hand, is robust and withstands extremely high-amplitude electromagnetic fields and can be used as a medium for compact THz devices \cite{Hamster14, Amico15, Leemans16, Ding17, Kumar18, Dechard19, Koulouklidis20, Kumar21, Kumar22, Liao23, Kim24}. Emerging ideas for plasma-based ultra-intense lasers \cite{Hur, Edwards, Vieux} are making laser-plasma methods attractive for future high-power THz sources.

Previous schemes for generating narrowband THz based on laser-plasma interactions have focused on utilizing electron plasma waves as sources of electromagnetic radiation. However, practical realisation of these radiators is challenging because of the inefficient coupling of sub-luminal plasma waves with electromagnetic waves in vacuum. Pukhov \textit{et al.} have recently suggested a solution to this problem by making the plasma wave super-luminal, which is achieved by exploiting the time-dependence of the plasma wavelength on a plasma density gradient \cite{Pukhov25}. 
Other plasma wave schemes, such as linear mode conversion \cite{Sheng26}, Cherenkov wakes \cite{Yoshii27}, and colliding wakefields \cite{Timofeev28} utilise plasma inhomogeneities and magnetization to achieve mode-coupling between the plasma and electromagnetic waves. These plasma waves are analogous to travelling wave antennae \cite{Melrose29}. In contrast, methods for generating stationary plasma \textit{oscillators} (in contrast to \textit{waves}) as multipolar radiators (\text{i.e.}, standing wave antennae), have rarely been investigated apart from the plasma dipole oscillator (PDO) excited by a colliding laser pulse \cite{Kwon30, Lee31}, which acts as dipole antennae. 

\begin{figure}[!t]
\centering
\includegraphics[width=0.5\textwidth]{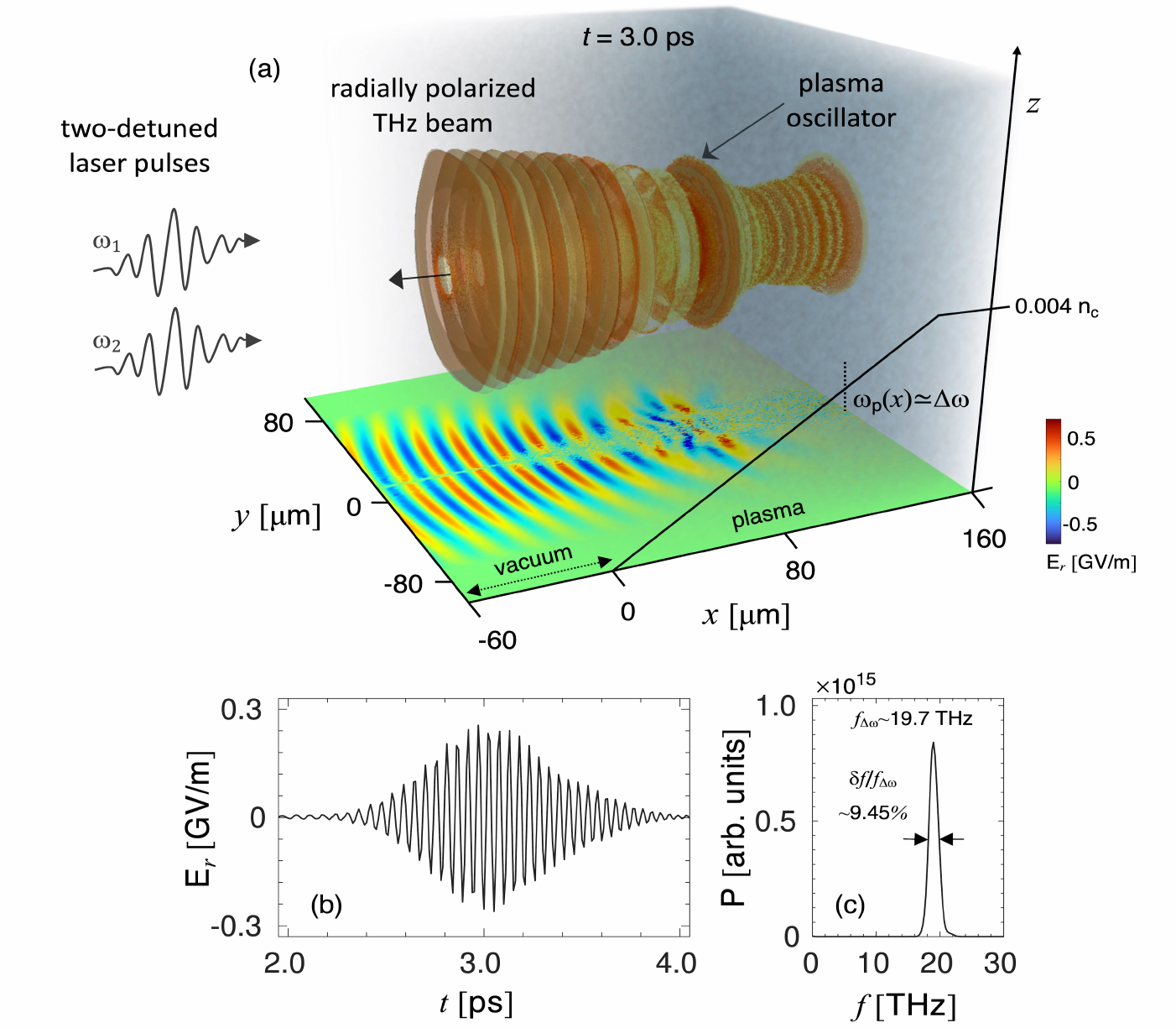}
\caption{(a) A 3-D volumetric contour plot and its 2-D slice snapshot (cut at $z=18$ $\mu$m, projected onto the bottom plane) depicting the radial component of emitted electric field at $t=3.0$ ps. The plasma density increases linearly from 0 to 0.004 $n_{c}$ over the range $x=0$ to 140 $\mu\text{m}$, then remains flat to 160 $\mu\text{m}$ (represented by a solid black line on the front plane); vacuum region ranges from $-60$ to 0 $\mu\text{m}$. The resonance point, where  $\omega_{p}(x)\cong\Delta\omega$, is indicated by a black dotted line. (b)-(c) Temporal profile of radiated field $E_{r}$ (GV/m) and its corresponding power spectrum determined at the vacuum side.}
\label{fig:fig1} 
\end{figure}

In this Letter, we introduce a novel idea for producing a plasma oscillator imbedded in a plasma slab with density gradient. The mechanism relies primarily on producing a localized plasma wave in a longitudinally narrow resonance region, where the local plasma frequency $\omega_{p}(x)$ matches the beat frequency $\Delta\omega$ of two detuned laser pulses, or a single bi-frequency pulse, that co-propagate up the density gradient. While the laser pulse drives electron motion in the resonant region, a plasma wave packet is generated, initially comprising several wavelengths. After the pulses have left the resonance region, the plasma wavenumber $k_{p}$ decreases with time because of the density gradient, which can be expressed in terms of the ray equation, $\partial k_{p}/\partial t = - \partial\omega_{p}/\partial x$, until it approaches zero (same mechanism is used in Ref. \cite{Pukhov25} to make the plasma wave superluminal, whereas, here we use it to obtain a localized oscillator). Over a certain duartion, the plasma wave packet $k_{p}\simeq 0$ and all electrons inside the packet oscillate in-phase. During this time, the transverse (radial) current, which is inherent in transversely finite plasma waves satisfying the Panofsky-Wenzel theorem \cite{Panofsky33}, also oscillates in-phase, forming a localized, radially-oscillating, plasma oscillator. The radial plasma oscillator acts as a planar antenna that emits an electromagnetic pulse  at THz frequencies, in the form of multipole radiation, but without requiring a mode conversion mechanism. The emitted radially polarized THz pulse propagates down the density gradient (backward) with high directionality, as shown in Fig.~\ref{fig:fig1}(a). 

The method proposed here has the advantage that it does not require fine adjustment of the plasma density to satisfy the resonance condition; the resonance point is always located somewhere on the density gradient. Furthermore, as the THz wave is emitted from a planar oscillator, its emission can be made highly directional (\text{i.e.}, with a small diffraction solid angle) by increasing the oscillator's transverse width, which is approximately the same as the laser pulse spot size, and the THz beam can be made useful for applications (\text{e.g.}, injection into the waveguide of a compact THz-band accelerator). This contrasts with the more divergent emission of other schemes \cite{Liao23, Kim24}, where the emission is generated from a line-like source (\text{e.g.}, a plasma wave train). The radial polarized fields can be readily converted to the fields with other polarizations when required for particular applications. 

We have used the fully relativistic three-dimensional (3-D), particle-in-cell (PIC) code EPOCH  \cite{Arber34} for the simulation study of the proposed radiator. We consider a bi-frequency driving laser pulse comprising two co-propagating and overlapping linearly-polarized laser pulses with frequencies $\omega_{1}=2\pi c/\lambda_{L1}$ and $\omega_{2}=2\pi c/\lambda_{L2}$ ($\lambda_{L1}=800$ nm and $\lambda_{L2}=760$ nm), respectively. 
Both laser pulses are chosen to have equal peak values of the normalized vector potential, $a=\text{max}(eE_{L}/m_{e}c\omega_{L})=0.1$, where $E_{L} \propto \exp[-r^{2}/r_0^{2}]\exp[-t^{2}/\tau_L^{2}]$, $\omega_{L}$, $m_{e}$, and $-e$ are the electric field, angular frequency, electron mass, and charge, respectively. The spot radius is $r_{0}=20$ $\mu\text{m}$ and pulse duration $\tau_{L}=0.3$ ps.
The laser pulse enters a plasma slab with a linear density ramp from 0 to 0.004 $n_{c}$ extending over 140 $\mu\text{m}$, where $n_{c}=m_{e}\varepsilon_{0}\omega_{1}^{2} /e^{2}$ is the critical density for $\omega_{1}$, and $\varepsilon_{0}$ is the free space permittivity. The resonance point is located at 97 $\mu\text{m}$ from the plasma-vacuum boundary, and the 3-D simulation domain is 220 $\mu$m, 160 $\mu\text{m}$, and 120 $\mu\text{m}$, in the $x$, $y$ and $z$ directions, respectively, with the grid sizes $dx=dy=dz=\lambda_{L1}/4$.

A radially polarized and highly directional THz beam, with a small diffraction angle, is emitted from a disk-shaped planar oscillator [Fig.~\ref{fig:fig1}(a)]. To determine the frequency and temporal profiles of the emitted THz pulse, we locate a point probe at $x=-$33.5 $\mu\text{m}$, $y=$ 32.5 $\mu\text{m}$, and $z=$ 18 $\mu\text{m}$. At this point, the temporal profile of radiation field includes several cycles with a peak field strength of $E_{r}\simeq 0.27$ GV/m [Fig.~\ref{fig:fig1}(b)]. Its power spectrum has a single dominant peak, centred at the beat frequency of the laser pulses, $f_{\Delta\omega}\approx19.7$ THz, with a relative spectral width of 9.45$\%$ [Fig.~\ref{fig:fig1}(c)], which can be made smaller by using a longer driver laser pulse to localize the oscillator more tightly.      

The characteristics of the emitted THz radiation described above have been determined as follows; for a linear density ramp, the ray equation gives $k_{p}=-(\Delta\omega/2L_{r})t+\Delta k$, where $L_{r}$ is the gradient scale length (for the linear density ramp used here, it is the distance from the position where the density is zero to that of the resonance point $x=x_{r}$), $\Delta\omega$ the beat frequency (\text{i.e.}, $\Delta\omega=\omega_{1}-\omega_{2}\cong\omega_{p}(x_{r})$) and $\Delta k=(k_{1}-k_{2})$ the beat wavenumber of the driving bi-frequency laser pulse (initial wavenumber of the plasma wave). The characteristic time for $k_{p}\simeq 0$, during which the wavenumber changes from $\Delta k/2$ to $-\Delta k/2$, is 
\begin{equation}
\centering
\tau\cong2L_{r}/c. 
\label{eq:one} 
\end{equation}
The ponderomotive force exerted on the plasma electrons produced by the bi-frequency laser pulse is given by,
\begin{equation}
\centering
{{f}}_{\text{PM}}=-\frac{m_{e}c^{2}\Delta k}{2}a_{1}a_{2}\left(\frac{i}{2}e^{i(\Delta kx-\Delta\omega t)} + c.c. \right),     
\label{eq:two} 
\end{equation}    
where $a_{1}$ and $a_{2}$ are the normalized vector potentials, \text{i.e.}, $a_{1}a_{2}=a^{2}e^{-2(t^2/\tau_{L}^{2})}$. The ponderomotive force drives a plasma wave that has an electric field $E_{x}(x,t)$\cite{Esarey35}, satisfying the equation of motion,
\begin{equation}
\centering
\left [ \frac{\partial^2}{\partial t^2} + \omega_{p}^{2}(x) \right]E_{x}(x, t)=F(x, t), 
\label{eq:three} 
\end{equation}
where $F(x, t)=-\frac{m_{e}c^{2}\Delta k\omega_p^{2}(x)a^{2}e^{-\frac{2t^{2}}{\tau_{L}^{2}}}}{4e}\left(i e^{i(\Delta kx-\Delta\omega t)} + c.c. \right)$. Using the Green's function of the harmonic operator, $G\left(t, t^{\prime} \right)=\omega_{p}^{-1}\sin{\left[\omega_{p}(t-t^{\prime})\right]}$ $(t>t^{\prime})$, $E_{x}$ in Eq.~(\ref{eq:three}), just after the driving pulses have passed, can be represented by, 
\begin{multline}
E_{x}\left(x, t\right)=\int_{-\infty}^{t}G\left(t, t^{\prime}\right)F\left(x, t^{\prime}\right)dt^{\prime} \\
\simeq\frac{\sqrt\pi m_{e}c^{2}}{4\sqrt2 e}\Delta k a^{2}\tau_{L}\omega_{p}(x)\exp{\left[-\frac{\left(\Delta\omega-\omega_{p}(x)\right)^{2}\tau_L^{2}}{8}\right]}\\
\times\left(\frac{1}{2}e^{i(\Delta kx-\Delta\omega t)} + c.c. \right).
\label{eq:four} 
\end{multline}
The transverse wakefield ($E_{\perp}$), which acts as an emitting plane antennae at THz frequencies, is related to the axial wakefield ($E_{x}$) by the Panofsky-Wenzel theorem, $\nabla_{\perp}E_{x}=(\partial/\partial\xi)E_{\perp}$, where $\nabla_{\perp}=\partial/\partial r$, $\xi=x-ct$. Notice that if $a=a_{0}\exp{{(-r}^{2}/r_{0}^{2})}$,
\begin{equation}
\centering 
E_{\perp}\sim\left[2r/(\Delta k r_0^2)\right]E_{x},
\label{eq:five} 
\end{equation}
 and $E_{\perp}$ has a maximum at $r=r_{0}/\sqrt{2}$.

\begin{figure}[!t]
\centering 
\includegraphics[width=0.5\textwidth]{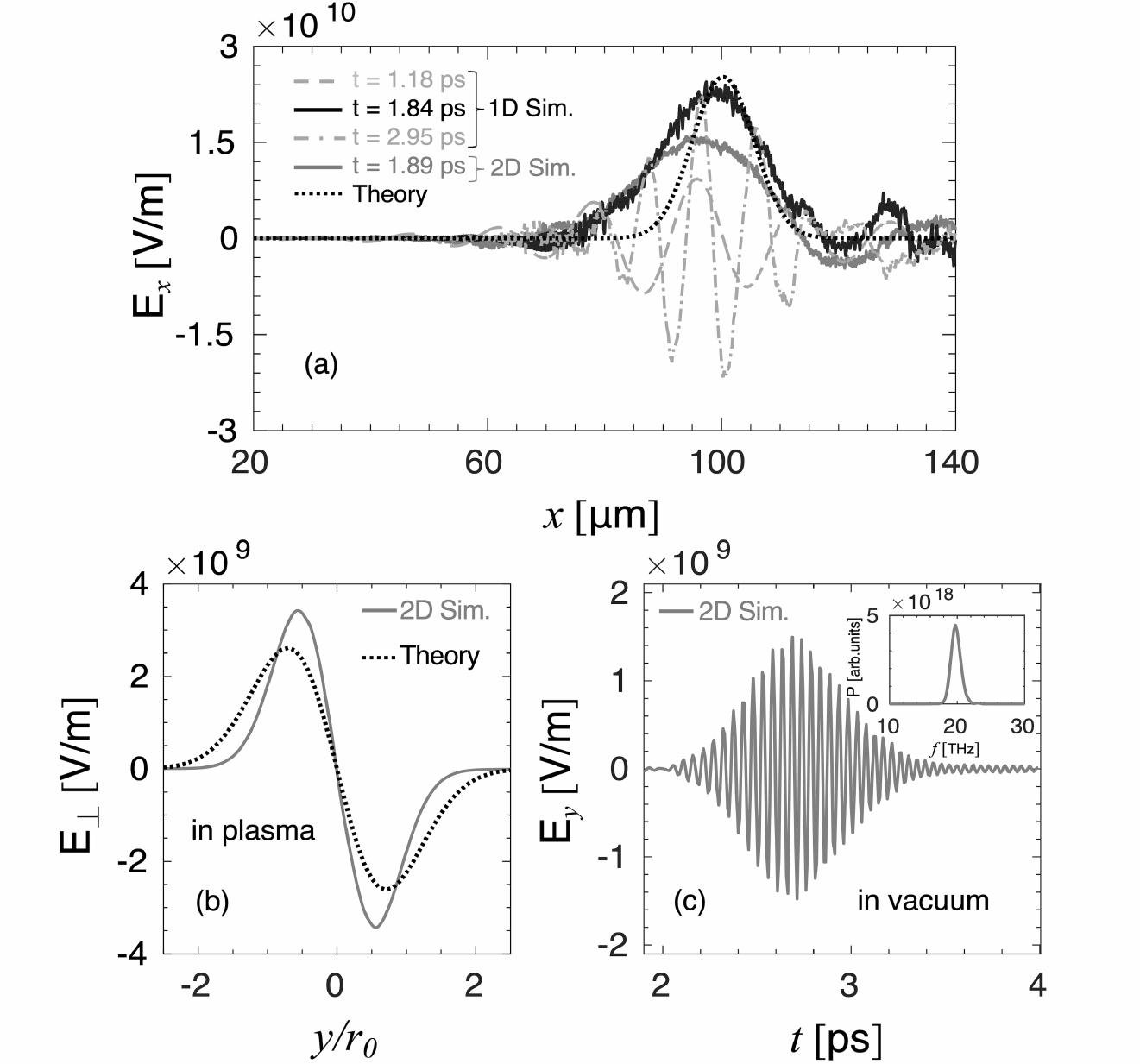}
\caption{(a) Longitudinal electric field $E_{x}$ at different times; 1-D results at $t=1.18$ ps (grey, dashed), $1.84$ ps (black, solid), $2.95$ ps (grey, dot-dashed) and 2-D at $t=1.89$ ps (grey, solid)  for the same laser and plasma parameters used for  Fig.~\ref{fig:fig1}. The dotted black line represents the theoretical profile of the plasma wave given by Eq.~(\ref{eq:four}). (b) Transverse electric field $E_{\perp}$ vs. $y/r_{0}$ at the resonance point, from 2-D simulations (grey, solid) and analytic theory (black, dotted). (c) Temporal evolution of THz field on vacuum side (at $x=-$33.5 $\mu\text{m}$ and $y=32.5$ $\mu\text{m}$) and its corresponding power spectrum (inset).}
\label{fig:fig2} 
\end{figure} 

We have performed 1-D and 2-D PIC simulations (because of limited computational resources) using the same set of parameters as for the 3-D simulation in Fig.~\ref{fig:fig1}. Figure~\ref{fig:fig2}(a) shows the spatial plot of the longitudinal electric field at times $t=1.18$ ps, $1.84$ ps, and $2.95$ ps (from 1-D simulations) and $t=1.89$ ps (from 2-D simulations). 
The plasma wave is excited by the ponderomotive force of the beat wave, which attains a very large amplitude in the resonance region (where $\omega_{p}(x)\cong\Delta\omega$). As time advances, the excited plasma wave begins to localize in that region. 
Initially, the plasma wave amplitude grows and reaches more than 20 GV/m in 1-D and slightly lower in 2-D simulations, which is close to the analytic amplitude given by Eq.~(\ref{eq:four}), but with a slightly wider localized oscillating region, compared with the theory. These plots show how the plasma wavelength varies with time in a density gradient plasma as both the oscillation frequency and wavelength are space-dependent. The emission from the plasma oscillator reaches a maximum around $t=1.8$ ps, when the localized plasma wave is contained in half a period. After this, the plasma wavenumber increases again, but with the opposite sign, and the THz wave train diminishes. The duration of the burst of THz radiation is given by Eq.~(\ref{eq:one}).
Figure~\ref{fig:fig2}(b) shows the transverse electric field near the resonance point (at $x\simeq97$ $\mu$m) at $t=1.84$ ps, obtained from the 2-D simulation (grey solid line) and analytic calculations (black dotted). The temporal evolution of the radiation field in vacuum obtained from 2-D PIC simulations is shown in Fig.~\ref{fig:fig2}(c) and the power spectrum (inset) is observed to have a single dominant peak at the central frequency $f_{\Delta\omega}\approx19.7$ THz. The peak field strength reaches up to 1.5 GV/m, which is higher than that from the 3-D PIC simulations with the same parameters mostly because the emission only diffracts in $y-$direction. However, the duration of the emitted pulse is the same as in 3-D PIC simulations. These results strongly indicate that the beat frequency mechanism will work efficiently in an experiment.

Here, we provide an analytic estimate of the important characteristics of radiation from the plasma oscillator. The emission field strength is proportional to the field strength of the oscillator, which is, from Eq.~(\ref{eq:four}), determined by the driving pulse energy per unit area (fluence), \text{i.e.}, $E_{THz}\propto{a}^{2}\tau_{L}$. The width ($\delta x$) of the local oscillating region of the plasma is determined by the exponential term in Eq.~(\ref{eq:four}). From $\left(\Delta\omega-\omega_{p}^{2}(x)\right)^2\tau_L^{2}/4\sim1$ and $n=n_{r}x/L_{r}$, where $n_{r}$ is the resonant density located at $x=L_{r}$, $\delta{x}\sim4L_{r}/{(\tau}_{L}\Delta\omega)$. The spectral width ($\delta\omega$) of the emission is directly related to $\delta x$, as the plasma electrons within $\delta x$ emit at their own local plasma frequency, leading to,
 \begin{equation}
\centering
\frac{\delta\omega}{|\Delta\omega|}\sim\frac{\delta x}{L_{r}}\sim\frac{4}{\tau_{L}\Delta\omega}. 
\label{eq:six} 
\end{equation}
As the THz emission is maximized when the emitting source width is smaller than the emission wavelength $\lambda_{THz}=2\pi{c}/\Delta\omega$, the condition for the most efficient emission is, 
\begin{equation}
\centering
\frac{2}{\pi}<\frac{c\tau_{L}}{L_{r}}.    
\label{eq:seven} 
\end{equation}\par

\begin{figure}[!t]
\centering
\includegraphics[width=0.44\textwidth]{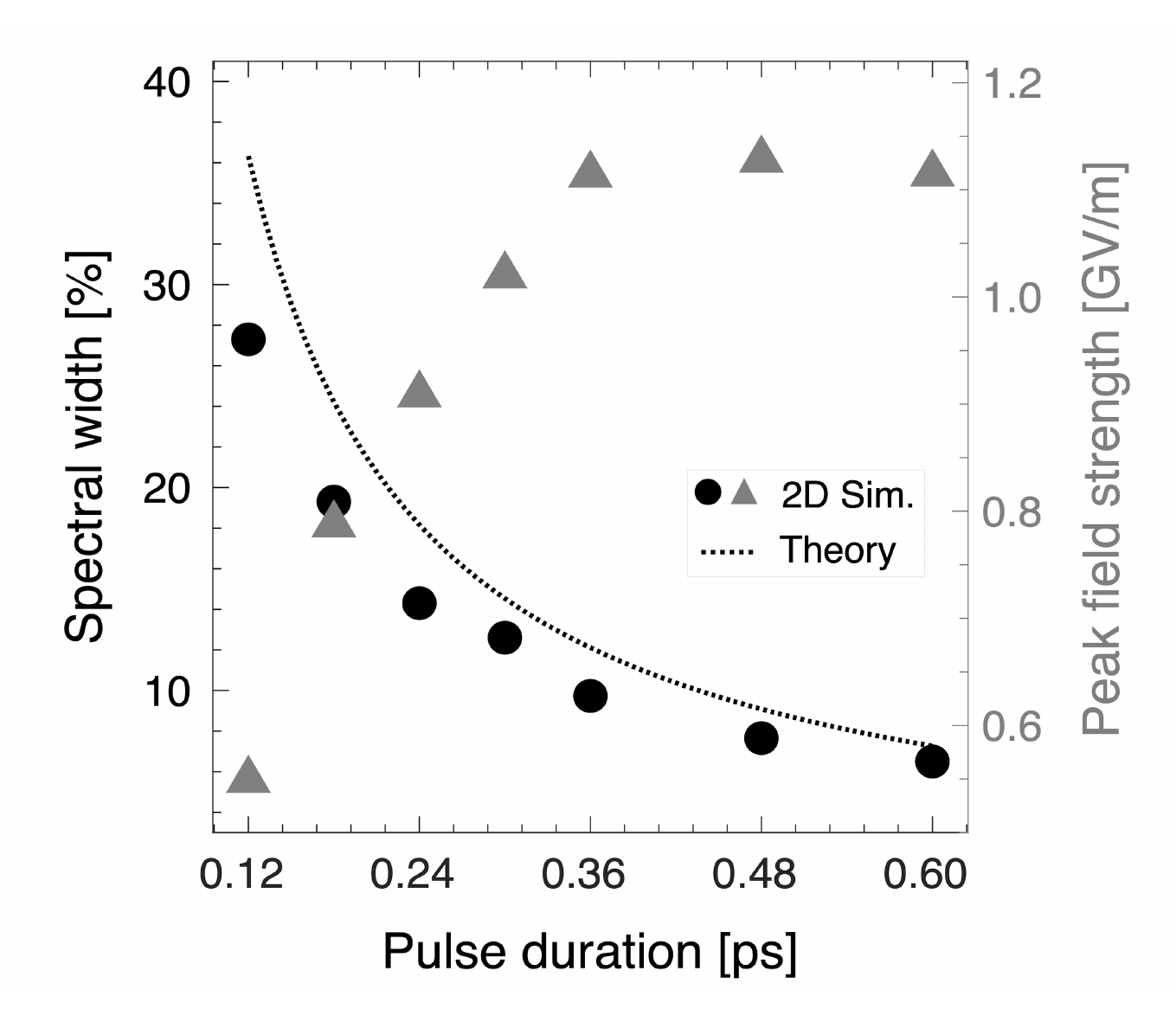}
\caption{Peak strength of emission electric field $E_{y}$ (triangles) and spectral width (circles) vs. duration of the driving laser pulses for beat frequency ${f}_{\Delta\omega}\approx14.6$ THz ($\lambda_{L1}=800$ nm and $\lambda_{L2}=770$ nm), and other parameters are the same as in Figs.~\ref{fig:fig1} and ~\ref{fig:fig2}. The dotted black line is the analytic spectral width given by Eq.~(\ref{eq:six}).}
\label{fig:fig3} 
\end{figure} 

Figure~\ref{fig:fig3} shows that the peak field strength of the emitted THz radiation (triangles), evaluated on the vacuum side at position $x=-$39.5 $\mu$m and $y=38.5$ $\mu$m, increases linearly with the driving pulse duration, as predicted by Eq.~(\ref{eq:four}), which is typical of resonantly driven oscillators. For pulse durations longer than 0.48 ps, it saturates due to the breaking of the resonance condition as the plasma wave becomes relativistic \cite{Liu36}. On the other hand, the spectral width (circles) decreases as the duration of the driving pulses increases. For comparison, the analytic spectral width [Eq.~(\ref{eq:six})] is shown as a black line.  A spectral width of about 6.6$\%$, for a duration of 0.6 ps is achieved in 2-D simulations. \par

\begin{figure}[!b]
\centering
\includegraphics[width=0.49\textwidth]{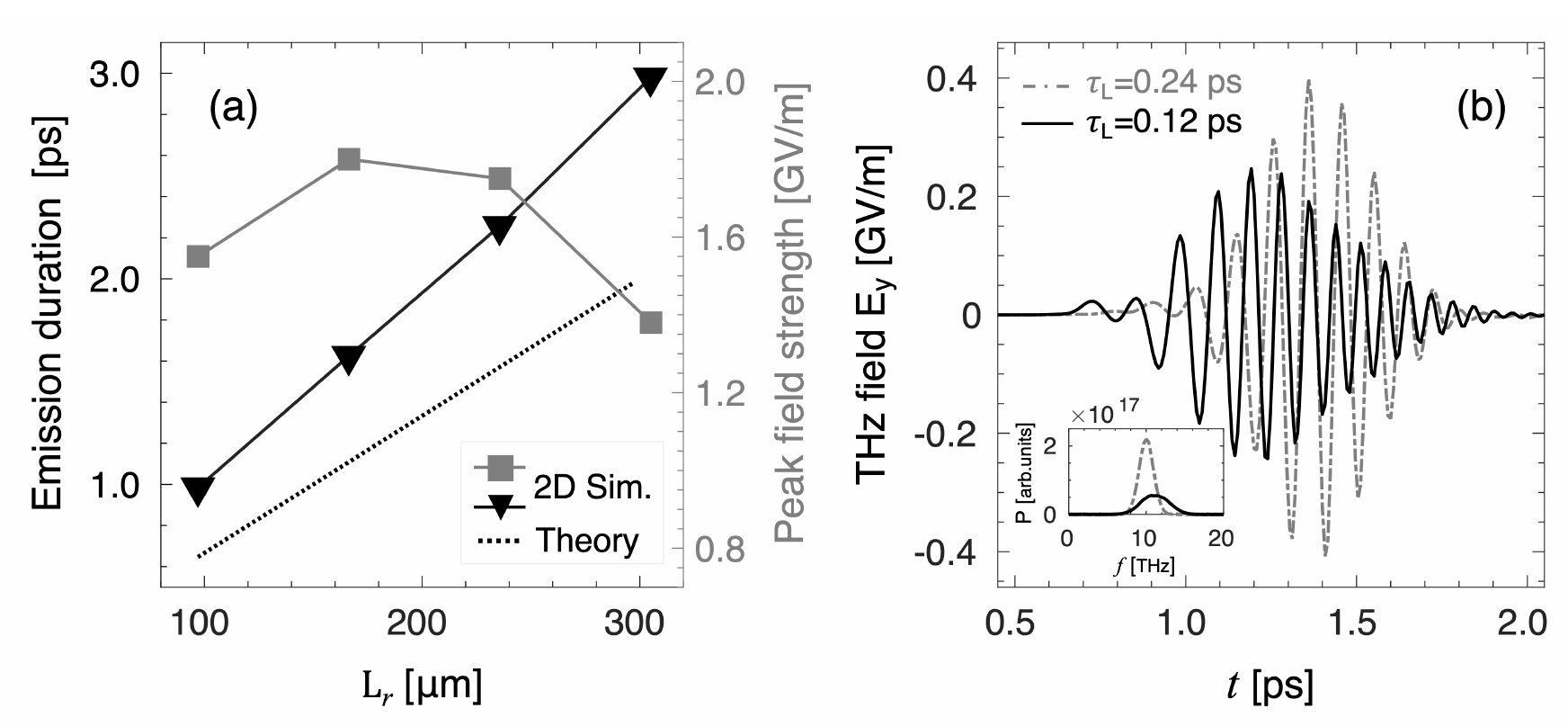}
\caption{(a) Duration of the THz burst (triangles) and peak field strength of THz (squares) vs. gradient scale length $L_{r}$ for parameters; ${f}_{\Delta\omega}\approx19.7$ THz, $\tau_{L}=0.24$ ps, $a=0.1$, $r_{0}=20$ $\mu\text{m}$, respectively. The dotted black line is the theoretical emission duration given by Eq.~(\ref{eq:one}).
(b) Comparison of temporal profiles of THz field $E_{y}$ and corresponding emission spectra (inset) for different driving pulse durations; $\tau_{L}=0.12$ ps (black, solid) and $0.24$ ps (grey, dot-dashed), ${f}_{\Delta\omega}\approx9.6$ THz and $L_{r}=97\mu$m.}
\label{fig:fig4} 
\end{figure}

Figure~\ref{fig:fig4}(a) shows the burst duration and peak field strength of THz radiation for gradient scale length varying from 97 $\mu\text{m}$ to 305 $\mu\text{m}$, for the case of $f_{\Delta\omega}\approx19.7$ THz. As predicted by  Eq.~(\ref{eq:one}), the duration of THz burst increases by a factor of 3 (triangles), which is the same factor of increase of $L_{r}$. The peak field strength of radiation (recorded in vacuum) depends relatively weakly on the density gradient (squares). This is consistent with the predictions of Eq.~(\ref{eq:five}), which does not depend on $L_{r}$. Also, as expected from Eq.~(\ref{eq:six}), there is no change in spectral width ($\delta{f}/{f}_{\Delta\omega}\simeq10$$\%$)  when $L_{r}$ is varied (not presented here).
When the driving pulse length is shorter than the gradient scale length $L_{r}$, the emitted radiation acquires a frequency-chirp [black solid in Fig.~\ref{fig:fig4}(b)]. The spectral bandwidth is larger and the spectral peak of the emission shifts towards higher frequencies, to 10.9 THz, which is higher than the beat frequency $f_{\Delta\omega}\approx9.6$ THz (dot-dashed line). This indicates that the plasma wave is not well localised, but is excited over a wide region, and therefore loses temporal coherence. As indicated by Eq.~(\ref{eq:seven}), the frequency-chirp in the emitted radiation can be reduced, and the full coherence can be obtained by using a pulse length longer than the gradient scale length. Also, note that the peak field strength of the THz emission is higher when its frequency is larger [$\sim1.5$ GV/m for 19.7 THz in Fig.~\ref{fig:fig4}(a), while $\sim0.4$ GV/m for 9.6 THz in Fig.~\ref{fig:fig4}(b)], as indicated by Eq.~(\ref{eq:four}) This also affects the efficiency of laser-THz conversion as will be shown below. 

We now calculate the energy conversion efficiency from driving laser pulse to emitted THz radiation. The plasma oscillator energy can be approximated by $U_{\text{PO}}\simeq\epsilon_{0}E_{r}^{2}c\tau_{\text{e}}A_{\text{f}}/2$, where $E_{r}$ is the transverse (radial) component of the oscillator's electric field, $A_{\text{f}}$ is its transverse area filling factor, and $\tau_{\text{e}}$ is the emission duration, which is approximately same as Eq.~(\ref{eq:one}). From Eq.~(\ref{eq:five}), with $\omega_{p}(x)\cong\Delta\omega$ and $r=r_{0}/\sqrt{2}$ (where $E_{r}$ is the maximum), except the factor $\epsilon_{0}/2$, 
\begin{equation}
\centering
 U_{\text{PO}}\simeq\frac{\pi\textbf{e}^{-1}}{8}\left(\frac{m_{e}c\Delta\omega}{e}\right)^{2}\left(\frac{a^{4}c^{2}\tau_{L}^{2}L_{r}}{r_{0}^{2}}\right)A_{\text{f}}.
 \label{eq:eight} 
\end{equation}                                                
Here \textbf{e} is the Euler number. The energy of driving pulse can be approximated by $U_{\text{L}}\simeq\left(m_ec\omega_{1}/e\right)^2a^2c\tau_{L}A_{\text{f}}$.  With $\zeta$ the conversion rate from oscillator energy to radiation energy (not all oscillator energy is emitted), the theoretical efficiency is then given by,
\begin{equation}
\centering
\eta=\frac{\zeta U_\text{PO}}{U_\text{L}}\simeq\zeta\frac{\pi\textbf{e}^{-1}}{8}\left(\frac{\Delta\omega}{\omega_{1}}\right)^{2}\frac{a^{2}c\tau_{L}L_{r}}{r_{0}^{2}} . 
 \label{eq:nine} 
\end{equation}    
With a conversion rate of approximately 0.5, the theoretically optimal efficiency is $\approx{0.01}\%$ [for the parameters used in Fig.~\ref{fig:fig1}]. Although this efficiency is low compared with broadband schemes such as CTR \cite{Liao23} or two-colour systems \cite{Kim24}, where typically the optimal efficiency is $\approx{0.1}\%$, it can be increased by using longer driving pulses, slower ramp up of density or by reducing the spot radius of laser pulse.

\begin{figure}[!t]
\includegraphics[width=0.39\textwidth]{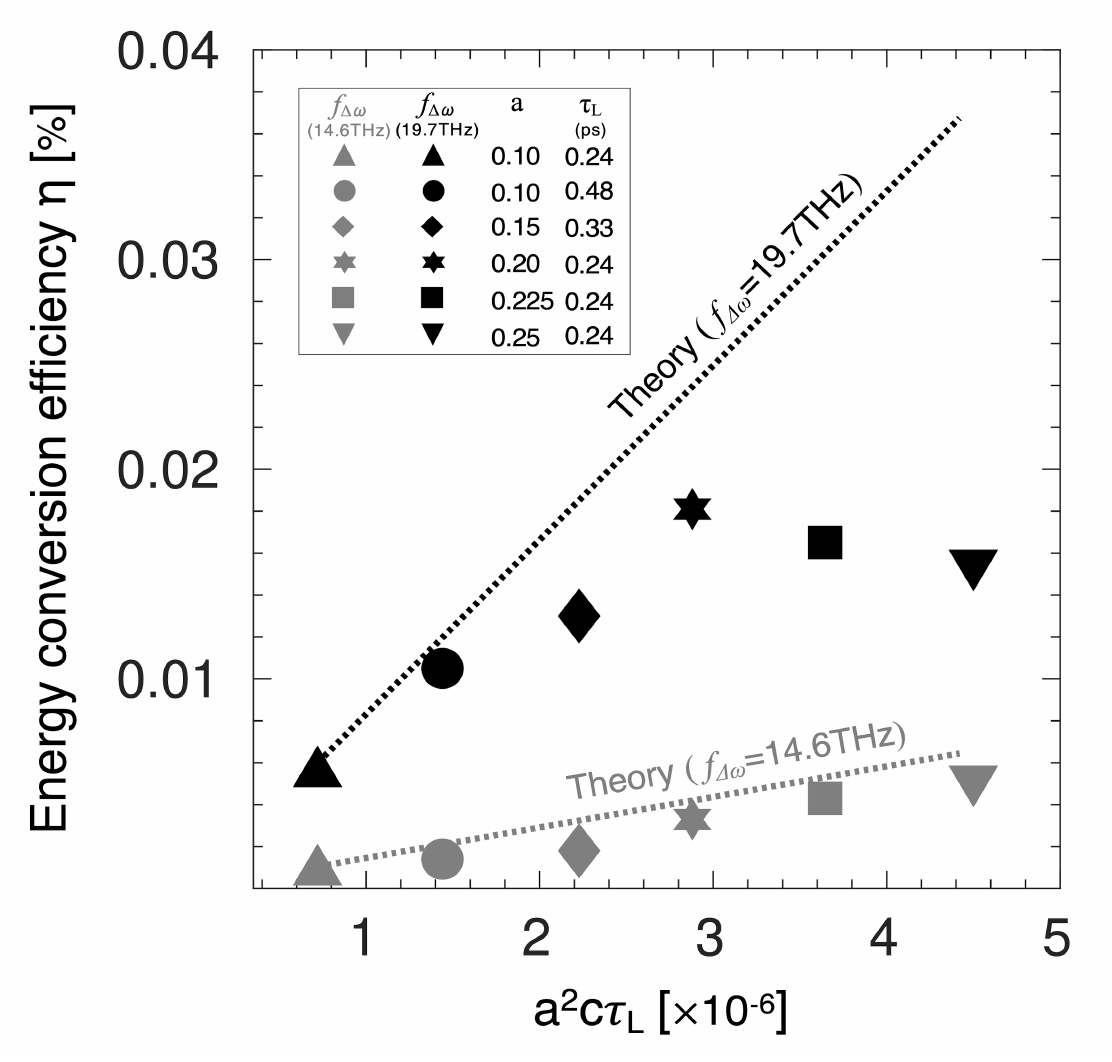}
\caption{Energy conversion efficiency from laser to THz vs. $a^{2}c\tau_{L}$ (laser fluence). Grey and black (upward-pointing triangle, circle, diamond, hexagon, square, and downward-pointing triangle) represent simulation results for different parameters of $f_{\Delta\omega}$, $a$, and $\tau_{L}$ with change of $L_{r}$ from 53.2 $\mu\text{m}$ to 166.3 $\mu\text{m}$. The dotted lines are theoretical curves given by Eq.~(\ref{eq:nine}).}
\label{fig:fig6} 
\end{figure} 

We compare the theoretical efficiencies with a set of 2-D simulations in Fig.~\ref{fig:fig6}. The total THz emission energy radiated per unit $z-$length (emission in $x-y$ plane) is given by, $dU_\text{rad}/dz={(\epsilon}_0/2)\int_{t_1}^{t_2}{cdt\int_{y_1}^{y_2}{E_{y}^{2}dy}}$. The theoretical efficiency is consistent with the simulation results [Fig.~\ref{fig:fig6}] and is shown to increase linearly with the driving laser fluence ($\sim a^{2}c\tau_{L}$) until it saturates due to nonlinear effects (described below). Furthermore, the increase in efficiency by a factor of 5.5 when $f_{\Delta\omega}$ changes from 14.6 to 19.7 THz, and $L_{r}$ from 53.2 $\mu$m to 166.3 $\mu$m (by comparing grey and black hexagons) is explained well by Eq.~(\ref{eq:nine}), \text{i.e.}, a factor of 1.8 from $(\Delta\omega/\omega_{1})^{2}$ and another factor of 3.1 from the change in $L_{r}$, leading to 5.6 times increase in efficiency. We observed that for large driving pulses intensities (roughly from $a\sim0.25$), the efficiency begins to saturate, likely due to relativistic effects and wavebreaking accompanied by thermalization of the plasma. It was also found that system becomes highly nonlinear for high-intensity driving pulses and begins to emit radiation at the beat frequency harmonics. Since harmonics are less shielded by plasma, they propagate not just in the backward direction but isotropically. However, the efficiency calculation consider only backward-propagating fundamental harmonic of the beat frequency.

In conclusion, we have presented a novel scheme for generating a radial plasma oscillator and emission of a narrowband THz pulse. To produce the plasma oscillator, a plasma wave packet is excited by the beat of a bi-frequency laser pulse, near resonance point on a plasma slab with a density gradient. The plasma wavenumber decreases with time because of density gradient, forming a transient, localized plasma oscillator, where all electrons inside the wave packet oscillate in-phase. The plasma oscillator emits an electromagnetic wave at the beat frequency without requiring a mode-conversion mechanism. This is a distinguishing feature of the proposed scheme, compared with conventional methods, where coupling a plasma wave with an electromagnetic wave is difficult. From multi-dimensional PIC simulations, we demonstrate the emission of radially-polarized, highly-directional THz waves with several GV/-level field strength and a spectral width less than 10$\%$. The novel scheme is readily applicable to compact THz-band accelerators or pump-probe experiments of materials.

\begin{acknowledgments}
This research work was supported by the National Research Foundation (NRF) of Korea (Grant Nos. NRF-2022R1I1A1A01055853, NRF-2020R1A2C1102236, NRF-2022R1A2C3013359, and NRF-2019R1C1C1003687). D.A.J. acknowledge support from the U.K. EPSRC (grant number EP/N028694/1) and received funding from the European Union’s Horizon 2020 research and innovation programme under grant agreement no. 871124 Laserlab-Europe.
\end{acknowledgments}

\end{document}